\documentclass{aa}  

\usepackage{graphicx}
\usepackage{txfonts}
\usepackage{hyperref}
\usepackage{lipsum}
\usepackage{booktabs}
\usepackage{subcaption}         
\usepackage{lscape}             
\usepackage{placeins}           

\usepackage{float}
\usepackage{xcolor}
\usepackage{flafter} 
\usepackage{txfonts}
\usepackage{natbib}

\usepackage[dvipsnames]{xcolor}
\definecolor{myblue}{HTML}{1F77B4}
\definecolor{mygreen}{HTML}{2CA02C}
\definecolor{m}{HTML}{D62728}
\definecolor{mymagenta}{HTML}{D33682}
\definecolor{codepurple}{HTML}{C42043}

\hypersetup{
unicode=true,           
colorlinks=true,        
citecolor=myblue,       
urlcolor=black        
}
\begin{document}

    \makeatletter
    \def\makeLineNumber{}
    \makeatother

   \title{\textsc{NOMAI}: A real-time photometric classifier for superluminous supernovae identification}
   \subtitle{A science module for the Fink broker}

   \author{E. Russeil\inst{1}\fnmsep\thanks{Corresponding author: etienne.russeil@astro.su.se}
   \and R. Lunnan\inst{1}
   \and J. Peloton\inst{2}
   \and S. Schulze\inst{3}
   \and P.~J. Pessi\inst{4}
   \and D. Perley\inst{5}
   \and J. Sollerman\inst{1}
   \and A. Gkini\inst{1}
   \and Y. Hu\inst{1}
   \and T.-W. Chen\inst{6}
   \and E. C. Bellm\inst{7}
   \and T. X. Chen\inst{8}
   \and B. Rusholme\inst{8}
   }

   \institute{Department of Astronomy, Oskar Klein Center, Stockholm University, SE-106 91 Stockholm, Sweden
   \and Université Paris-Saclay, CNRS/IN2P3, IJCLab, Orsay, France
   \and Department of Particle Physics and Astrophysics, Weizmann Institute of Science, 234 Herzl St, 76100 Rehovot, Israel
   \and Astrophysics Division, National Centre for Nuclear Research, Pasteura 7, 02-093 Warsaw, Poland
   \and Astrophysics Research Institute, Liverpool John Moores University, 146 Brownlow Hill, Liverpool L3 5RF, UK
   \and Graduate Institute of Astronomy, National Central University, 300 Jhongda Road, 32001 Jhongli, Taiwan
   \and DIRAC Institute, Department of Astronomy, University of Washington, 3910 15th Avenue NE, Seattle, WA 98195, USA
   \and IPAC, California Institute of Technology, 1200 E. California Blvd, Pasadena, CA 91125, USA
   }

\abstract{

Superluminous supernovae (SLSNe) are one of the most luminous stellar explosions known, yet they remain poorly understood. Because they are intrinsically rare, efficiently identifying them in the large alert streams produced by modern time-domain surveys is essential for building larger observational samples and enabling spectroscopic follow-up. We present \textsc{NOMAI}, a machine learning classifier designed to identify SLSN candidates directly from photometric alerts in the ZTF stream, using light curves accumulated over at least 30 days. It does not require any spectroscopic redshift and is running in real time within the Fink broker. ZTF light curves are transformed into a set of physically motivated features derived primarily from model-fitting procedures using SALT2 and Rainbow, a blackbody-based multi-band fitting framework. These features are used to train an XGBoost classifier on a curated dataset of labeled ZTF sources constructed using literature samples of SLSNe, along with TNS and internal ZTF labeled sources. The final training dataset contains 5280 unique sources, including 225 spectroscopically classified SLSNe. On the training sample, the classifier reaches 66\% completeness and 58\% purity. Deployed within the Fink broker, \textsc{NOMAI} has been running continuously since 18/12/2025 on the ZTF alert stream and publicly reports SLSN candidates every night by automatically posting them to dedicated communication channels. Based on this, we also report the first two-month as an evaluation period, where the classifier successfully recovered 22 of the 24 active SLSNe reported on the Transient Name Server. The achieved performances, particularly the high completeness in real-time operations, demonstrate that the classifier provides a valuable tool for experts to efficiently scan the alert stream and identify promising candidates. In the near future, \textsc{NOMAI} is intended to be adapted to operate on the Legacy Survey of Space and Time conducted by the Vera C. Rubin Observatory, which is expected to uncover an unprecedented number of transients, making machine learning based photometric classification essential.

}

\keywords{supernovae: general, methods: data analysis, techniques: photometric, surveys, astronomical databases: miscellaneous, methods: numerical}

\maketitle

\section{Introduction}
\label{sec:intro}

Superluminous supernovae (SLSNe) are rare stellar explosions whose peak luminosities exceed those of standard SNe by one to two orders of magnitude \citep{Quimby2011,GalYam2012,slsn}. Since their identification as a distinct population \citep{Quimby2011}, SLSNe have attracted considerable attention due to their extreme energetics. Although they display very diverse photometric properties, their light curves are typically evolving more slowly than those of normal SNe and often display blue colors \citep[e.g.,][]{Lunnan2018,Angus2019,DeCia2018,Chen_2023,Gomez2024,pessi_SLSNII}. Because of their high luminosities, they are detectable at large cosmological distances, making them valuable probes for the high-redshift universe.

Despite significant observational and theoretical progress over the past decade, the physical mechanisms powering these explosions remain debated. Several scenarios have been proposed, including magnetar spin-down \citep{Ostriker1971,Kasen2010,Woosley2010,Vurm2021}, interaction with dense circumstellar material \citep{Chatzopoulos2012,Sorokina2016,Wheeler2017,Chen2023b}, long-term fallback accretion of material onto a black hole \citep{Dexter2013,Moriya2018}, or pair-instability explosions \citep{pisn,Barkat1967,Rakavy1967,Woosley2002,Heger2002}. Further constraining these models requires larger samples of well-observed events spanning a broad range of environments and redshifts. However, SLSNe are intrinsically rare \citep[e.g.,][]{Prajs2017,Perley2020,SLSN_rates} and only represent a small fraction of the overall supernova population. Identifying them efficiently in wide-field surveys is therefore a key step toward improving our understanding of these extreme transients.

Modern time-domain surveys are now producing unprecedented volumes of transient alerts. The Zwicky Transient Facility \citep[ZTF, ][]{ztf1, ztf2, ztf3, Masci} currently delivers hundreds of thousands of alerts per night, and the recent Vera C. Rubin Observatory will increase this number to ten million through the Legacy Survey of Space and Time \citep[LSST, ][]{lsst}. In this context, the classical approach of spectroscopic follow-up for every candidate becomes impossible. Only a small fraction of transient discoveries can realistically be spectroscopically classified, making automated photometric classification an essential part of modern astronomy. The alert streams produced by modern surveys require dedicated infrastructures capable of processing and distributing transient classifications in real time. Alert brokers \citep{AMPEL, Fink, Sanchez2021, antares, lasair, BABAMUL}\footnote{\url{https://pitt-broker.readthedocs.io/en/latest/\#}} have emerged to fulfill this role by ingesting survey alerts, enriching them with additional information, and providing value-added products to the community. In this context, machine-learning classifiers can naturally be integrated within alert brokers to operate directly on the incoming data stream, and multiple works have explored the specific task of classifying SLSN light curves, although some classification pipelines are still developed and evaluated outside broker infrastructures.

Some approaches rely on multi-class classifiers in which SLSNe constitute one of several transient classes, such as the ALeRCE data broker \citep{Sanchez2021}, or Superphot+ \citep{superphot+} running on the ANTARES broker \citep{antares}, both of which nightly classifies alerts from the ZTF data stream. Other pipelines are specifically designed to identify SLSN events among the broader population of transients. The "NEural Engine for Discovering Luminous Events" \citep[NEEDLE,][]{needle}, hosted within the Lasair broker \citep{lasair}, uses a hybrid neural network to process both the photometry and the telescope images to classify SLSN and TDE sources. "Finding Luminous and Exotic Extragalactic Transients" \citep[FLEET, ][]{fleet, fleet2} attempts to classify SLSNe by relying on host galaxy association and photometric feature extraction to train a random forest classifier. Finally, by combining SuperRAENN \citep{superraen}, a recurrent autoencoder architecture, with Superphot \citep{superphot}, a versatile feature-extraction pipeline based on light-curve model fitting, \cite{SuperRAENN_SLSN} developed a classifier designed to identify SLSNe in Pan-STARRS1 data \citep{PS1}. Furthermore, other works have focused on anticipating the LSST era using simulated datasets such as the Photometric LSST Astronomical Time-series Classification Challenge \citep[PLAsTiCC,][]{plasticc} to train SLSN classification models \citep{SCONE, panchroma_plasticc, TLW}. Similarly, more recent works using the Extended LSST Astronomical Time-Series Classification Challenge \citep[ELAsTiCC,][]{elasticc} have explored analogous problems using a more realistic LSST-like data stream format \citep{fraga2024, ATAT}.

Among the brokers, Fink \citep{Fink} has been operating on the ZTF public stream since late 2019, while also helping prepare for the upcoming Rubin era. It provides an infrastructure that allows the scientific community to deploy dedicated “science modules”, which operate on the alert stream to compute additional features and classifications. This framework makes it possible to run machine-learning models continuously on the incoming alerts and distribute their outputs in near real time.

In this work, we present \textsc{NOMAI}, a machine-learning classifier designed to identify SLSN candidates directly from ZTF photometric alerts and deploy it within the Fink broker as a science module. The classifier relies on physically motivated features extracted from transient light curves without requiring a spectroscopic redshift, and is optimized to operate in a real-time streaming environment. Its integration within the broker enables the automatic identification and public sharing of SLSN candidates from the ZTF alert stream. In what follows, we describe the methodology used to build the classifier, in particular the construction of an informative dataset (Section \ref{sec:data}),  the extraction of the features, and the training of the model (Section \ref{sec:methodo}). We then present the performance obtained on the training data (Section \ref{sec:train_results}), as well as the practical metrics achieved on the ZTF stream during the first months of operation (Section \ref{sec:fink}). Finally, we discuss the model's limitations and its performance within the current landscape (Section \ref{sec:discussion}), before concluding (Section \ref{sec:conclusion}).

\section{Training dataset}
\label{sec:data}

In this section, we describe the construction of the training dataset. Because the ZTF alert stream contains a large fraction of non-transient sources, an initial filtering step is required before performing any feature extraction or classification. We therefore apply a sequence of selection criteria to isolate likely transient events and ensure sufficient photometric quality for the analysis. We then describe the origin of the data and the procedure used to assign labels to the sources.

\subsection{Transient cut}
\label{subsec:transient}

The real-time ZTF data stream processed by the Fink broker consists of compact data packets called alerts \citep{ZTF}, generated whenever a source exhibits a significant brightness variation relative to the reference template image. Hence, a single transient source will trigger multiple alerts. Photometry-wise, they do not contain the full light curve generated by the events, but only their last detection along with 30 days of history. However, given the long-lived nature of SLSNe, it is crucial that the complete light curve is used to provide the most informed classification possible. Consequently, the Fink API must be queried nightly to retrieve the complete public photometric history of each source. But this operation is computationally expensive, and given the large stream of alerts provided by ZTF, which averages $\sim 250,000$ alerts per night, it is infeasible to perform this step on all alerts every night. Therefore, simple cuts must be applied in order to remove as much non-transient contamination as possible upstream. They constitute highly computationally efficient solutions that can be applied to process even the largest datasets. 

Hence, the \emph{transient\_complete} alert filter was integrated to the Fink broker. It is based on the Bright Transient Survey filtering pipeline\footnote{A public implementation can be found here: \url{https://zenodo.org/records/4054129}. It is available on Fritz under the name Redshift Completeness Factor (RCF).} \citep{rcf_deep_1, rcf_deep_2}, which enables high transient completeness while considerably reducing contamination. The full \emph{transient\_complete} pipeline consists of 8 conditions:

\begin{itemize}
    \item[$\medcirc$] not \emph{faint}: \emph{magpsf} (PSF-fit difference magnitude) is currently brighter than 19.8, or the source had a very recent detection brighter than 19.
    \item[$\medcirc$] no \emph{pointunderneath}: is likely not sitting on top of or blended with a star in Pan-STARRS.
    \item[$\medcirc$] \emph{positivesubtraction}: is brighter than the template image.
    \item[$\medcirc$] \emph{real}: is likely a genuine astrophysical transient and not an artifact.
    \item[$\medcirc$] \emph{stationary}: is not a moving source.
    \item[$\medcirc$] no \emph{brightstar}: is likely not contaminated by a nearby bright star.
    \item[$\medcirc$] not \emph{variablesource}: is likely not a variable star. 
    \item[$\medcirc$] not \emph{roid}: is likely not a solar system object
\end{itemize}

The exact implementation is publicly available on Fink\footnote{\url{https://github.com/astrolabsoftware/fink-filters/blob/master/fink_filters/ztf/filter_transient_complete/filter.py}}. On average, this filter alone removes more than 99\% of the alert stream, reducing it to a few hundred alerts per night and making it possible to retrieve the photometric history for the remaining events.

\subsection{Quality cuts}
\label{subsec:quality}

The performance of any alert-classification system strongly depends on the minimal amount of information it requires from a light curve to produce a reliable output. Because SLSNe are generally slowly evolving events, it is possible to accumulate a substantial number of observations before classifying a source, thereby improving the purity. However, since one of the primary goals of the classifier is to direct scientists toward promising candidates for spectroscopic follow-up, the classification should ideally be produced while the source is still bright and young. We balance early detection and improved purity by imposing minimal requirements on the photometry. To ensure that a substantial portion of the transient’s evolution is sampled, we require that each source has at least 30 days of photometric history before we can provide a classification. Given the characteristic timescale of SLSNe, this still allows the classifier to return a prediction near peak brightness for sources first detected close to explosion; however, faster-evolving SLSNe may be missed due to this requirement (Section \ref{subsec:lim_and_future}). On top of a duration criterion, we also impose that the sources have more than 7 photometric observations in total, and more than 3 observations in each particular filter (\textit{ZTF-g} and \textit{ZTF-r}). These requirements guarantee sufficiently constrained fits during the feature-extraction stage (Section \ref{subsec:extraction}).

\subsection{Data source}
\label{subsec:data_source}

Our training dataset is built around the labels available on the Fritz\footnote{\url{https://fritz.science/}} platform. Fritz is an extension of Kowalski\footnote{\url{https://github.com/skyportal/kowalski}} and SkyPortal \citep{skyportal, skyportal2}, serving as an alert broker, multi-survey data archive, marshal, and tool for target management and follow-up. It also allows the creation of working groups, that receive a sub-stream of alerts based on customizable filters. Over the years, experts across multiple areas of transient astronomy have used this interface to assign spectroscopic classifications to transient sources. Using these high-quality labels for our dataset, the machine learning classifier will learn from numerous confidently labeled events. We proceed by collecting the \emph{objectId} of every source that passed the filter presented in Section \ref{subsec:transient} and \ref{subsec:quality}.

As Fritz is a private platform not fully accessible to the broader scientific community, we instead rely on the public Fink broker to retrieve the corresponding public light-curve data. Hence, the list of \emph{objectId} is passed to the Fink API to collect all associated ZTF light curves (\textit{ZTF-r} and \textit{ZTF-g} filters only), including detections ($SNR>5$) and bad quality points ($SNR>3$) but excluding upper-limits. In order to gather as many positive events as possible, we complement this dataset by adding potentially missing SLSN sources reported in the most recent and complete sample papers. Hence, we gather all available ZTF SLSN-I light curves reported by \citet{gomez_SLSNI}, which regroups all SLSN-I from previous sample papers and even provides a python package to access the data\footnote{\url{https://github.com/gmzsebastian/SLSNe}}. We also add all SLSN-II reported by \citet{pessi_SLSNII}. Fritz automatically filters out most known AGN and stellar contaminants. However, excluding these sources removes light-curve behaviors that would otherwise be encountered in real-time classification and could lead to incorrect predictions by the classifier. To account for this, we manually include $\sim200$ such non-transient sources that we found could be misclassified as SLSNe during preliminary tests of NOMAI on the real-time alert stream. These contaminants are mainly non-variable stars produced by faulty difference imaging, high–proper-motion stars, and AGNs with only a few detections spread over several years. Such events may appear either as flat light curves or as very slowly evolving ones, and can therefore be mistaken for SLSNe.

The final intent of NOMAI is to provide real-time classification of transients, and therefore the light curves that will be processed will rarely display a complete photometric picture of the sources. Indeed, classification is performed as soon as the previously stated criteria are met, even if the source is barely fading, is at peak, or is even still rising. Therefore, the training set should be representative of this reality and learn all the possible light curve configurations. Each light curve is therefore decomposed into multiple pseudo-alerts by iteratively removing the most recent observation, thereby generating truncated light curves that represent progressively earlier evolutionary stages of the transient. Following this method, a transient with 100 data points can be split into 100 pseudo-alerts (among which a fraction would not pass the quality cuts).  We name them pseudo-alerts since they encompass the real-time nature of alerts, but they are not restricted by their 30 days history structure.

\subsection{Labeling}
\label{subsec:labeling}

After applying the quality cuts presented in Sections \ref{subsec:quality} and \ref{subsec:transient}, we are left with 8849 unique sources. We then proceed to label them by combining Fritz and TNS labels, which are not always simultaneously available and do not necessarily match. Given that we use a classification scheme from three distinct sources (Fritz, TNS, and sample papers), the attribution of a definitive label is not always trivial. We proceed using the following priorities:

\begin{enumerate}
    \item If there exists a label from a published SLSN sample study, we prioritize it. Given that such analysis proposes highly curated samples, we consider this as the most reliable source. 
    \item If only a TNS or a Fritz label is available, or if both labels are available and in agreement, we use that label.
    \item If both TNS and Fritz labels are available but are in minor disagreement, we use the closest common denomination as a final label (e.g, if one is SLSN-I and the other is SLSN-II, the final label will be SLSN).
    \item If both TNS and Fritz labels are available and are in major disagreement (e.g, TDE and SLSN-I), the object is labeled as “Ambiguous”
    \item Finally, if no label is available, the source is marked as unclassified. 
\end{enumerate}

In addition to this labeling procedure, we must account for the fact that some SLSNe may be missed by the original classifications, assuming a peak magnitude definition. Following the method proposed by \cite{pessi_SLSNII}, we set an absolute magnitude threshold of $-19.9$ in any passband, below which a source is labeled as superluminous \citep[note that other similar threshold values could also be used, such as][using $-20$]{Gomez_2022}. Based on this limit, we further refine the labels by gathering from Fritz all available spectroscopic redshifts in order to compute the peak absolute magnitude of the light curves.  We compute the distance modulus using the \texttt{astropy.cosmology} package \citep{astropy:2022} with cosmological parameters coming from the Planck 2015 results \citep{planck2015}: $H_0=67.8$ $\mathrm{km\,s^{-1}\,Mpc^{-1}}$, $\Omega_0=0.308$, $\Omega_\Lambda=0.692$. Following the procedure from \citet{Chen_2023}, we use only the cosmological term $-2.5\times \log(1+z)$ for the K-correction \citep{Hogg2002}. Finally we correct for the Milky Way extinction using the \texttt{dust\_extinction} python package \citep{Gordon2024} assuming a \citet{F99} reddening law with $R_v = 3.1$. 

\begin{figure*}
    \centering
    \includegraphics[width=.9\linewidth]{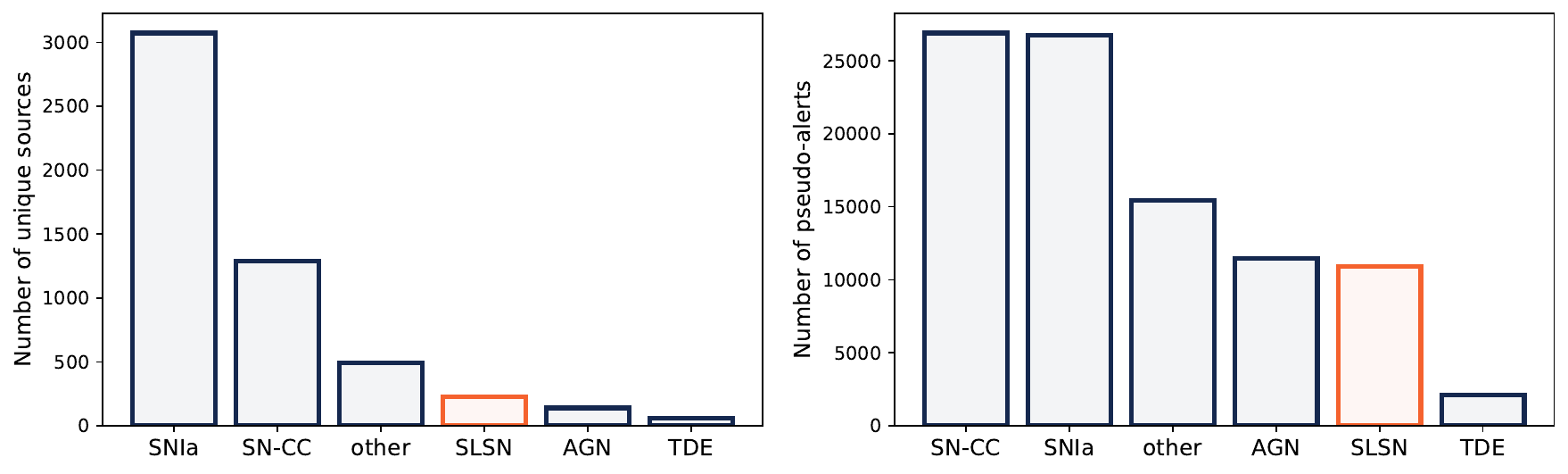}
    \caption{Class distribution of the training dataset for the unique sources (left panel) and for the pseudo-alerts generated from them (right panel). The ``other'' class mostly includes variable stars, novae, and the non-transient contaminants added.}
    \label{fig:class-distrib}
\end{figure*}

Any object brighter than the threshold that was previously labeled as core-collapse supernova is upgraded to the SLSN category. In many cases this procedure is essential. For example, SLSN-II sources are often initially labeled as SN IIn based on spectroscopic classification, but the labels are not always updated even when experts recognize that the source is photometrically superluminous. In the end, after applying all cuts and labeling procedures, we obtain a definitive label for 5280 unique ZTF sources, among which 225 unique SLSNe. The left panel of Figure \ref{fig:class-distrib} shows the class distribution of the different types of sources. It shows the high imbalance of the dataset with SLSNe representing $\sim4\%$ of all sources, with a large majority of SNe Ia or core-collapse supernovae. The right panel shows the dataset after converting the full light curves to pseudo-alerts (Section \ref{subsec:data_source}). Since each individual source produces multiple pseudo-alerts, the total number of events increases by roughly a factor of twenty. However, longer-duration transients, such as SLSNe, generate more pseudo-alerts than shorter-lived events. It leads to an improvement regarding the dataset imbalance, with SLSN alerts representing $\sim12\%$ of all pseudo-alerts. \\

\section{Methodology}
\label{sec:methodo}

In this section, we describe the methodology used to construct the classification pipeline. Because machine-learning algorithms require homogeneous inputs, the irregular and noisy nature of astronomical light curves must first be translated into a set of numerical features. Therefore, we begin by describing the feature-extraction procedure applied to each pseudo-alert, before presenting the implementation and optimization of the machine-learning classifier used to identify SLSN candidates.

\subsection{Feature extraction}
\label{subsec:extraction}

Light curves are noisy, multidimensional, and unevenly sampled objects, making the task to learn their patterns challenging, given that most machine learning algorithms require homogeneous input for training. Although some neural network architectures such as Recurrent Neural Networks \citep[e.g,][]{SuperNNova,RNN_goto} or transformers \citep[e.g,][]{ATAT,oher_transformer} can handle sequential data without major transformation, most light curve classification pipelines have relied on the extraction of physical features \citep[e.g,][]{ishida2018,fleet,Sanchez2021,leoni2022,fink_tde}. This method relies on the automatic computation of predefined set of numerical values carrying physical information able to separate different types of transients, effectively converting inhomogeneous light curves into fixed sized arrays. The extraction should be constructed such that as much information as possible about the transient light curve is encompassed into a small amount of features.

An effective way to characterize a transient light curve is to fit it with an appropriate mathematical model and use the optimized parameters of that fit as compact, descriptive features. Following this principle, each light curve is fitted using the Rainbow framework \citep{rainbow}, as implemented in the \texttt{light-curve} Python package\footnote{\url{https://github.com/light-curve/light-curve-python}}. This approach performs a single simultaneous surface fit across all fluxes and passbands, rather than treating them independently. Features extracted using the Rainbow method have been shown to outperform the standard approach in classification tasks, particularly for SLSNe \citep{rainbow}, and it has already been adopted in several classification pipelines \citep{fraga2024, fink_tde, rainbow_used_by_random_people}.

Rainbow relies on the assumption that the transient emission can be approximated by a blackbody, thereby introducing temperature parameters to be fitted in addition to the standard flux-related parameters. Therefore, two models should be chosen, one to describe the bolometric flux evolution and one to describe the temperature evolution through time. We choose to model the bolometric evolution using the Bazin \citep{bazin} function, defined as:

\begin{equation}
    F(t;\ t_0, A, \tau_\mathrm{rise}, \tau_\mathrm{fall}) = A \times \frac{e^{\frac{-(t-t_{0})}{\tau_\mathrm{fall}}}}{1+e^{\frac{(t-t_{0})}{\tau_\mathrm{rise}}}},
    \label{eq:bazin}
\end{equation}

where $t_0$ is a reference time parameter, $A$ is the amplitude, $\tau_{rise}$ is the characteristic rise time, and $\tau_{fall}$ is the characteristic decay time. The model describes the flux evolution as an exponential rise followed by an exponential decay. It enables the proper description of most transients, but is insufficient to characterize more complex behaviors such as double peaks or a plateau phase during the decay. However, preliminary tests with more complex functions (e.g., additional shape parameters) did not improve the performance of the classifier, while increasing the computational cost.

We model the temperature using a decaying sigmoid function. It provides a first-order approximation of the temperature evolution of a supernova, whose effective temperature decreases with time as the ejecta expand and cool. It is expressed as:

\begin{equation}
T(t; t_0, T_{min}, T_{max}, t_{color}) = T_{min} + \frac{(T_{max} - T_{min})}{1 + \exp{\frac{(t-t_{0})}{t_{color}}}},
\label{eq:temp} 
\end{equation}

where $t_0$ is a reference time parameter, $T_{max}$ is the initial temperature, $T_{min}$ is the final temperature, and $t_{color}$ is the exponent driving the speed of the cool down phase. Following \citet{rainbow}, we assume a common $t_0$ parameter for both the flux and temperature models, thereby reducing the number of free parameters by one. Combining the flux and the temperature component, we obtain a final model consisting of 7 parameters that we can extract as features. The \texttt{light-curve} package used to perform the fit also provides uncertainties associated with each optimized parameter. Hence, we compute additional signal-to-noise ratio features by dividing each best-fit parameter by its corresponding uncertainty. Finally, we extract the $\chi^2$ of the fit, representing how well the model matches the data.\\

Given the high number of SNe Ia in the training set, we also perform a SALT2 \citep{salt2} fit using the \texttt{sncosmo}\footnote{\url{https://sncosmo.readthedocs.io/en/stable/index.html}} python package. SALT2 is an empirical parametric model designed to describe Type Ia supernovae, which are known for their relative homogeneity. The model is characterized by three physical parameters: $x_0$ an amplitude parameter, $x_1$, the light-curve stretch, and $c$ the color. In addition, \texttt{sncosmo} can also treat the redshift $z$ as a free parameter which gets simultaneously fitted. Finally, the  $\chi^2$ of the fit is also collected and constitutes an additional feature from the SALT2 model. \\

In addition to the model-fitting procedures, we compute a set of features directly from the raw light curves, or from the alert packet. These features capture valuable information about the data while remaining computationally inexpensive to derive. A summary of all these additional features is provided in Table \ref{tab:stat_features}. Most of them are meant to provide additional description of the data. The features $q15$ and $q85$ are intended to identify and mitigate cases where a recent transient is associated with a single spurious observation from several years ago—causing it to appear artificially long-lived—or, conversely, where an older transient produces a single detection long after its main event. The $distnr$ constitute one way for the model to separate SLSNe from Tidal Disruption Events (TDEs) which occur at the center of their host galaxy. 

\begin{table}[]
    \centering
    \renewcommand{\arraystretch}{1.2}
    \begin{tabular}{l p{6cm}}
         \hline
         \textbf{Features} & \textbf{Description}  \\
         \hline
         
         $flux\_amplitude$ & Difference between the $max$ and $min$ flux. \\
         $max\_slope$ & Maximum flux slope between two consecutive observations.\\
         $std\_flux$ &  Standard deviation of the flux. \\
         $duration$ & Time difference between the last and the first observation. \\
         $q15$ & 15th percentile of the observation epochs.\\
         $q85$ & 85th percentile of the observation epochs.\\
         $skew$ &  Skewness of the flux. Measures the asymmetry of a distribution. \\
         $distnr$ &  Mean angular distance to the nearest object in the reference image. Available in ZTF alert packets \\
         $E(B-V)$ & Milky Way extinction. Retrieved using \texttt{SkyCoord} from the \texttt{astropy} package.\\
         \hline
         
    \end{tabular}
    \caption{Summary of complementary features computed without data-fitting procedure.}
    \label{tab:stat_features}
\end{table}

We derive statistical features, apply both Rainbow and SALT2 fitting procedures, and extract the physically meaningful parameters from these models for each alert. Altogether, this process yields a set of 27 features per alert, as detailed in Appendix (Table \ref{tab:all_features}). Figure \ref{fig:parameter_space} displays a projected view of the parameter space for pairs of the 8 most discriminative features (Figure \ref{fig:feature_importance}). It clearly illustrates that SLSNe occupy a subregion of the parameter space and are distributed differently than the other sources. This suggests that the extracted features are physically informative and that a machine-learning classifier can be trained to effectively separate the classes.

\begin{figure}[htbp]
    \centering
    
    \begin{subfigure}{0.24\textwidth}
        \centering
        \includegraphics[width=\linewidth]{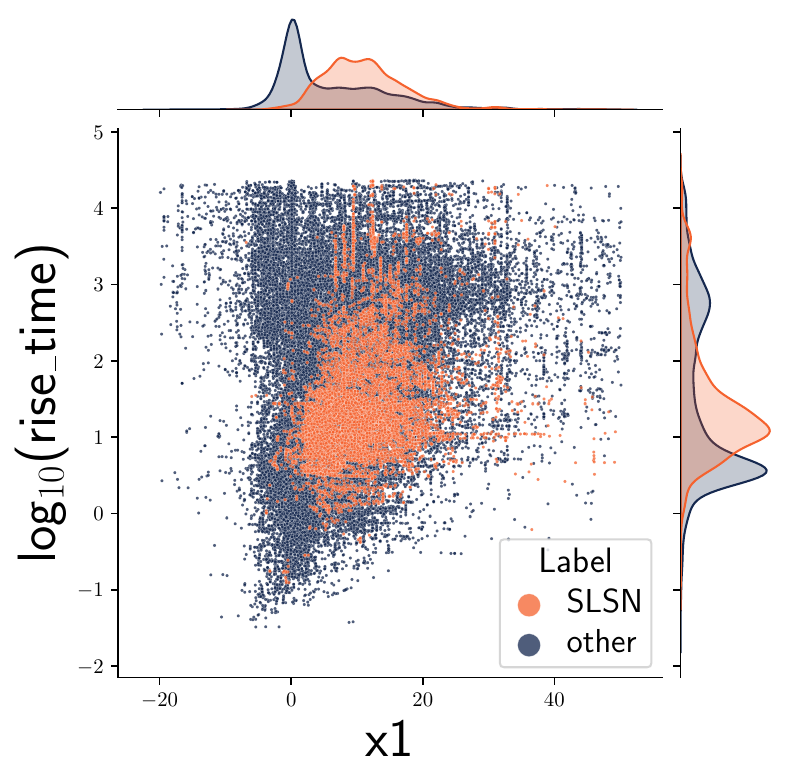}
    \end{subfigure}
    \hspace{0.00\textwidth}
    \begin{subfigure}{0.24\textwidth}
        \centering
        \includegraphics[width=\linewidth]{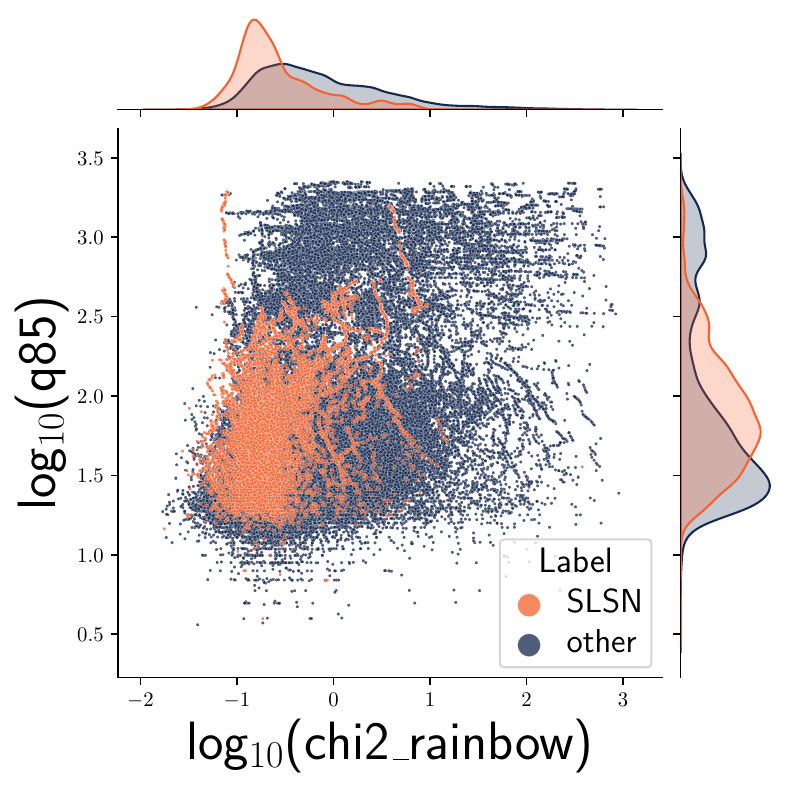}
    \end{subfigure}
    \vspace{0.00\textwidth}
    \begin{subfigure}{0.24\textwidth}
        \centering
        \includegraphics[width=\linewidth]{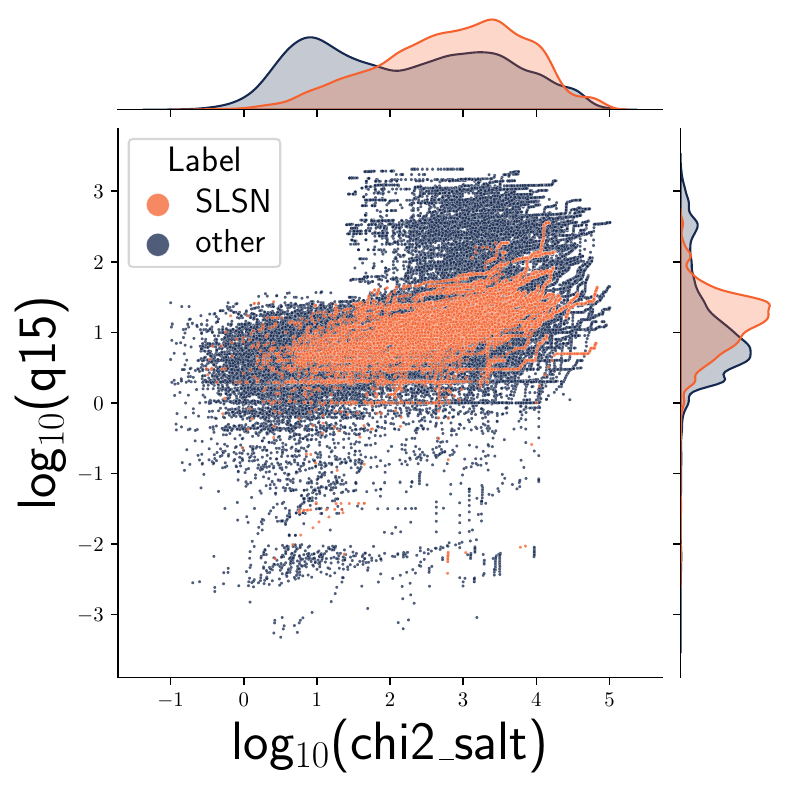}
    \end{subfigure}
    \hspace{0.00\textwidth}
    \begin{subfigure}{0.24\textwidth}
        \centering
        \includegraphics[width=\linewidth]{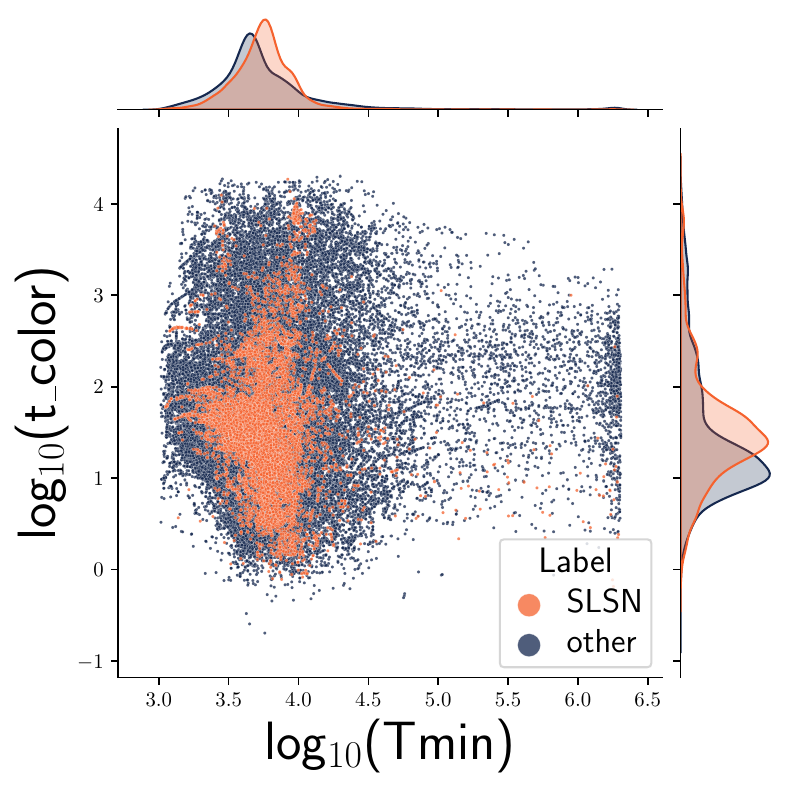}
    \end{subfigure}
    
    \caption{Projection of the parameter space along 4 pairs of the 8 most important features (Figure \ref{fig:feature_importance}). for SLSNe (orange) and the other alerts (blue). Normalized kernel density estimated feature (KDE) distributions are also displayed.}
    \label{fig:parameter_space}
\end{figure}

\subsection{Classifier implementation}
\label{subsec:extraction}

We chose to train an XGBoost\footnote{\url{https://xgboost.readthedocs.io/}} \citep{XGBoost} classifier to separate SLSNe from the other sources. It is a highly efficient and scalable machine learning algorithm that combines multiple weak learners, implemented as decision trees, to form a stronger, more accurate model. To mitigate dataset imbalance, we weighted each label inversely to its frequency, so that SLSNe have a similar impact on the learning process as the other sources. 

The output probabilities of the classifier are subsequently refined using an additional physically motivated criterion based on the absolute magnitude of the sources, computed from the publicly available photometric redshifts. This step aims to leverage external information that is readily accessible for a large fraction of sources in wide-field surveys, even for newly discovered transients.

Therefore, for each sources, we query the Sloan Digital Sky Survey \citep[SDSS,][]{SDSS} for a photometric redshift, $z$, and its associated uncertainty $\Delta z$. If no photometric redshift is available, the probability is left unchanged. Otherwise, an upper limit of the peak absolute magnitude (corresponding to a lower limit on the absolute brightness) is computed following the methodology of Section \ref{subsec:labeling} and using $z-\Delta z$ as a redshift. If the upper limit of the Milky Way corrected absolute magnitude computed from the photometric redshift is less bright than $-19.75$~mag, the probability of the source being a SLSN is set to 0. Note that this threshold is independent from the $-19.9$~mag used for the labeling procedure. It is a purely empirical number that we find to be effective at cutting a significant amount of false positive without losing any true SLSN. 

We then perform an hyperparameter optimization procedure to select the most effective model. We perform 1000 training iterations, sampling the XGBoost hyperparameters from a random grid, and we select the set that maximizes the F1-score defined as:

\begin{equation}
F_1 = 2 \times \frac{\mathrm{purity} \times \mathrm{completeness}}{\mathrm{purity} + \mathrm{completeness}}
\label{eq:f1}
\end{equation}

The hyperparameters optimized are $reg\_alpha$ the L1 regularization term, $reg\_lambda$ the L2 regularization term, $max\_depth$ the maximum depth of a tree, $max\_delta\_step$ the maximum delta step for each tree's weight, and the learning rate. The precise sampling distributions and the final optimal values are presented in Table \ref{tab:hyperparam_grid}. 

\begin{table}[htbp]
    \centering
    \renewcommand{\arraystretch}{1.2}
    \begin{tabular}{p{2.4cm} p{2.9cm} p{2.2cm} }
        \hline
        \textbf{Hyperparameter} & \textbf{Description} & \textbf{Optimal value} \\
        \hline
        $reg\_alpha$  & Uniform(0, 100) & 68.3 \\
        $reg\_lambda$ & Uniform(0, 10) & 0.34\\
        $max\_depth$ &  UniformInt(3, 5) & 5\\
        $max\_delta\_step$ & UniformInt(0, 2) & 2\\
        $learning\_rate$ & Uniform(0.01, 0.31) & 0.26\\
        $n\_estimators$ & Fixed & 100\\

        \hline
    \end{tabular}
    \caption{Summary of the hyperparameter search space used for the random grid search. The right column shows the optimized values used to train the final model.}
    \label{tab:hyperparam_grid}
\end{table}

The number of trees, $n\_estimators$ has been manually set to 100, as increasing the value does not lead to any meaningful improvement but reduces the computation performances. We also emphasize that the $max\_depth$ sampling distribution is intentionally kept to small values in order to improve the generalization performances of the final classifier. Since the SLSN class is not yet fully characterized, we favor simpler and more general models to reduce the risk of excluding atypical or previously unseen behaviors that could still correspond to SLSNe.

\section{Training results}
\label{sec:train_results}

\begin{figure}
    \centering
    \includegraphics[width=1\linewidth]{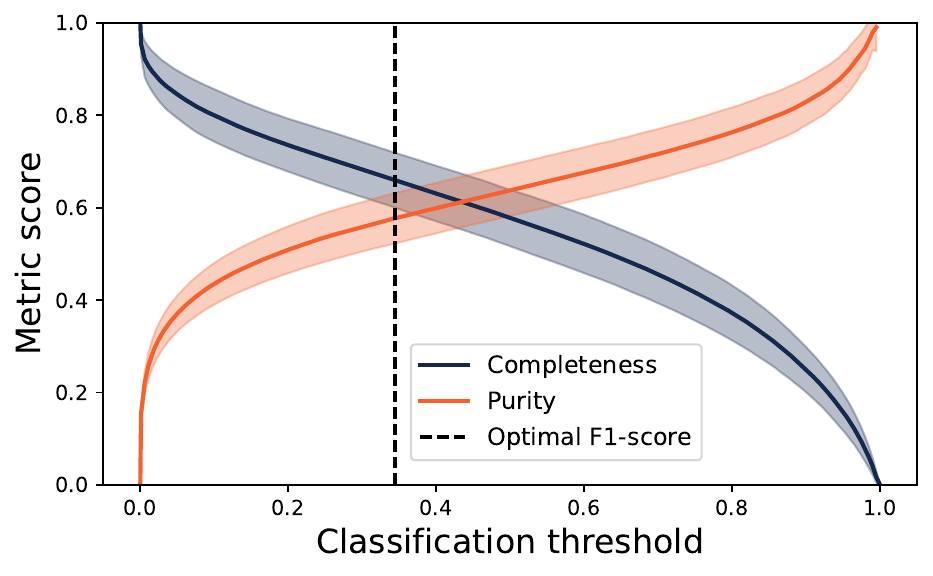}
    \caption{Evolution of the purity and completeness scores as a function of the classification threshold. The uncertainties correspond to the standard deviation of the bootstrapping iterations.}
    \label{fig:threshold}
\end{figure}

We assess the performances of the model by performing 100 iteration of bootstrapping \citep{bootstrapping}, which provides more stable results than K-folding methods given the data imbalance. The random sampling of the procedure is performed on unique objects, such that alerts from a given source may only be present in either the training or the testing sample, thereby preventing information leak. Figure \ref{fig:threshold} displays the evolution of the scoring metrics depending on the classification threshold. As it increases from 0.1 to 0.9, the completeness decreases from 80 to 25\%, whereas the purity increases from 44 to 83\%, illustrating the trade-off between sensitivity and contamination. A threshold value of $\sim 0.35$ maximizes the final F1-score (Equation \ref{eq:f1}), therefore any alert receiving a score higher than this is classified as a SLSN. 

\begin{figure}
    \centering
    \includegraphics[width=1\linewidth]{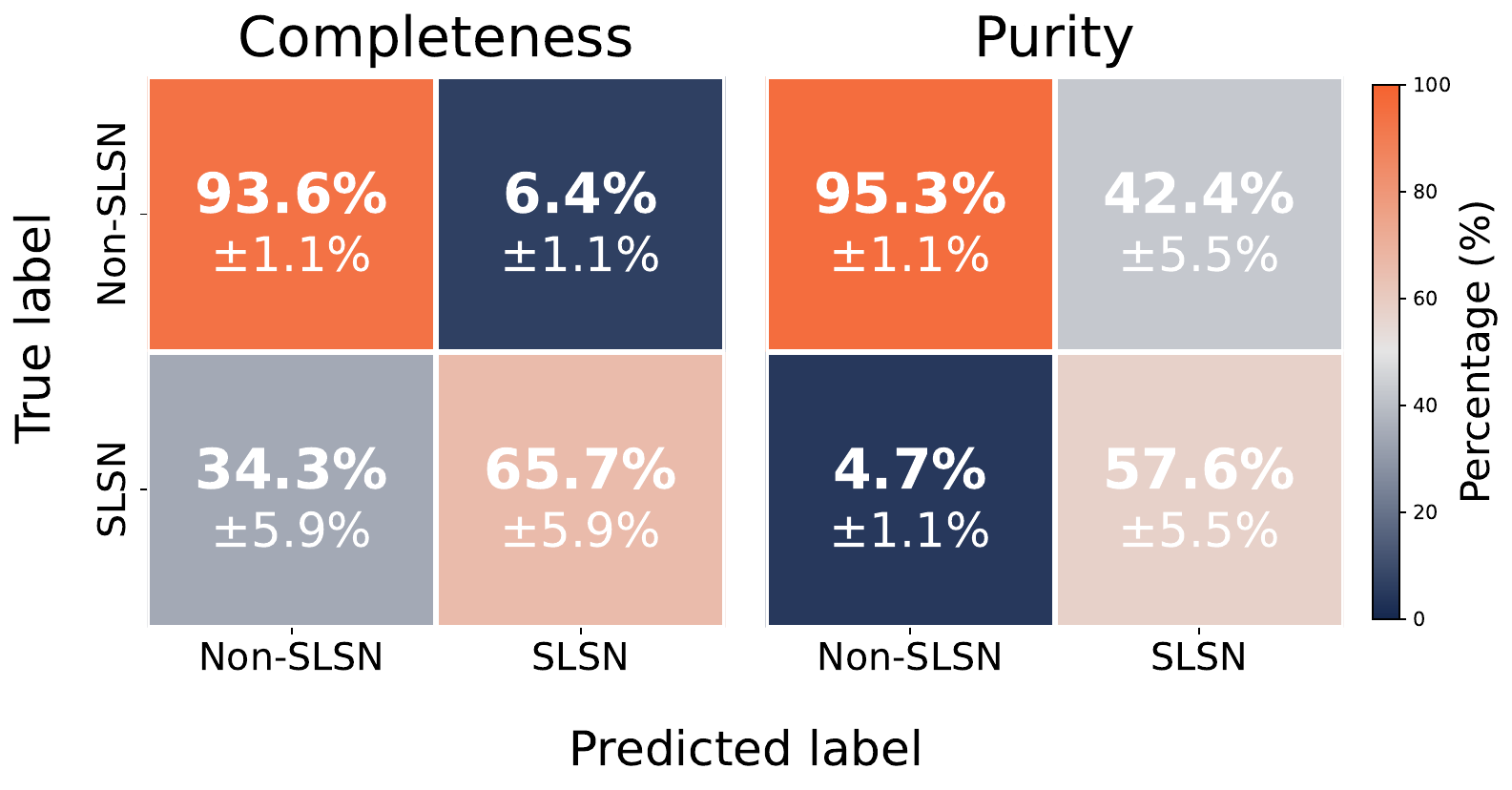}
    \caption{Confusion matrices of 100 iteration of bootstrapping. The left (right) matrix is normalized on completeness (purity). The main value of each cell corresponds to the mean performance of the bootstrapping procedure, while the value displayed below corresponds to the standard deviation.}
    \label{fig:confusion}
\end{figure}

Figure \ref{fig:confusion} presents the completeness and purity performances of \textsc{NOMAI} using the optimal threshold. Despite the high imbalance, it reaches 58\% purity and 66\% completeness in the task of classifying SLSNe. Identifying two out of three SLSNe using only light curves, and without access to spectroscopic redshifts, represents a significant achievement.

This is particularly noteworthy given that the distinction between SLSNe-II and SNe IIn remains debated, with both classes potentially forming a continuum rather than being intrinsically distinct \citep{SNIIn_continuum}. This is reflected in the false positives, since approximately one third of the misclassifications belong to the SN IIn class. Another third arise from SNe II, which are known for their extended plateau phase during decline and can therefore mimic the characteristically slow evolution of SLSNe. The remaining major source of contamination is AGNs, which account for 13\% of the alerts incorrectly classified as SLSNe. Their stochastic variability can randomly produce luminous, slowly evolving light curves resembling those of SLSNe. Although the pre-filtering procedure (Section \ref{subsec:transient}) removes most AGN contamination, it is statistically unavoidable that some brightening AGN sources will temporarily behave like transients and pass the selection cuts.

\begin{figure}
    \centering
    \includegraphics[width=1\linewidth]{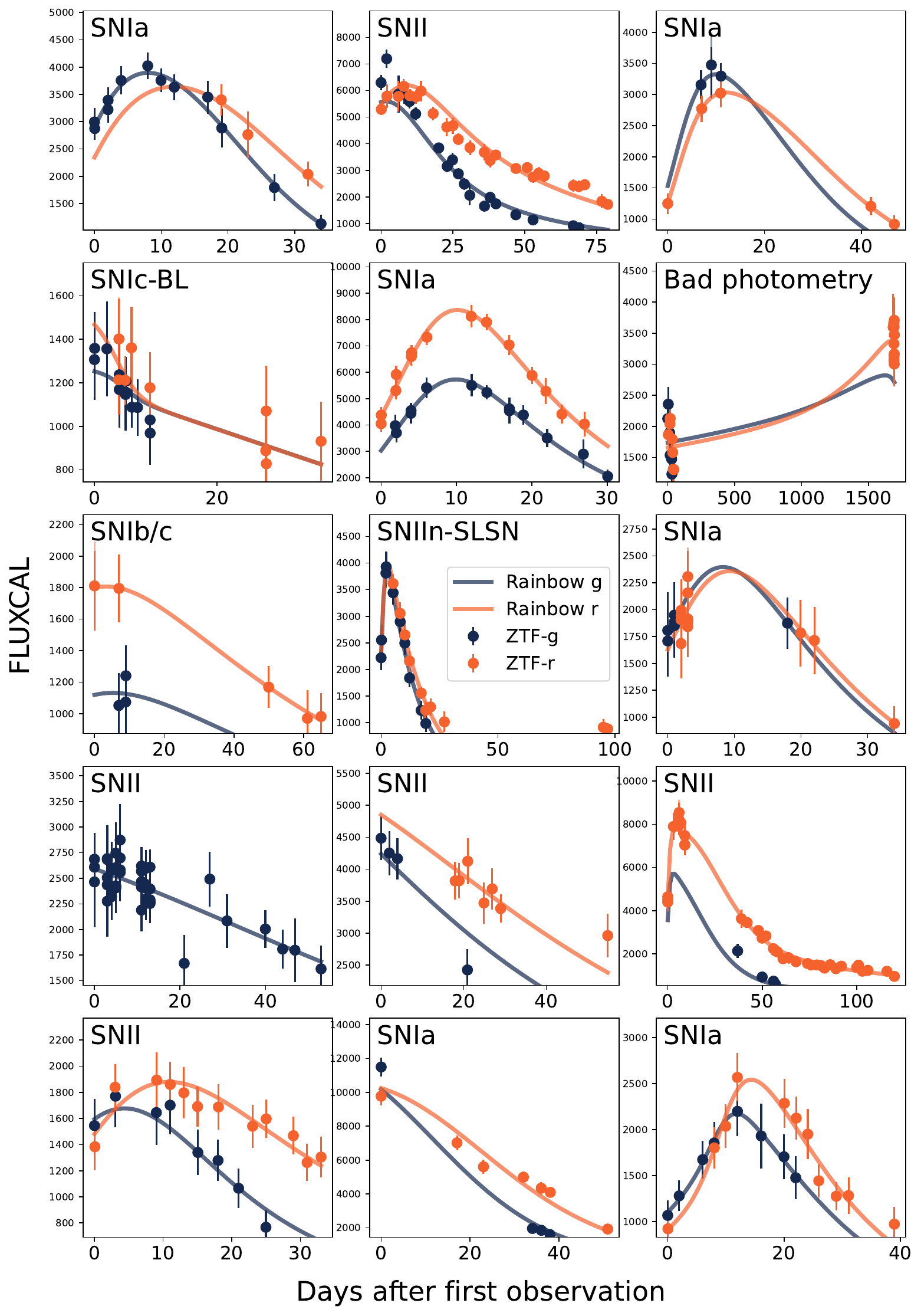}
    \caption{Collection of the 15 objects with a score closest to the classification threshold. The Rainbow model fitted to the data is also displayed.}
    \label{fig:border_alerts}
\end{figure}

In order to better understand the limitations of the classifier, it is useful to study alerts with a score close to the classification threshold, which highlights events for which the classifier is the least confident. Figure \ref{fig:border_alerts} presents the 15 objects closest to the threshold. The only SLSN located at the decision boundary is a relatively fast evolving but luminous SN IIn, that was reclassified as an SLSN IIn during the labeling procedure (Section \ref{subsec:labeling}). Besides this event, approximately a third of the borderline cases are SNe Ia, partly reflecting their prevalence in the dataset (Figure \ref{fig:class-distrib}). However, they also correspond to atypical SNe Ia, characterized by poorly sampled light curves, high Milky Way extinction, a missing rising phase, or luminosities at the upper end of the SN Ia distribution. The remaining events near the decision threshold are predominantly slow-evolving core-collapse supernovae, in particular SNe II ($1/3$ of the events).

\begin{figure}
    \centering
    \includegraphics[width=1\linewidth]{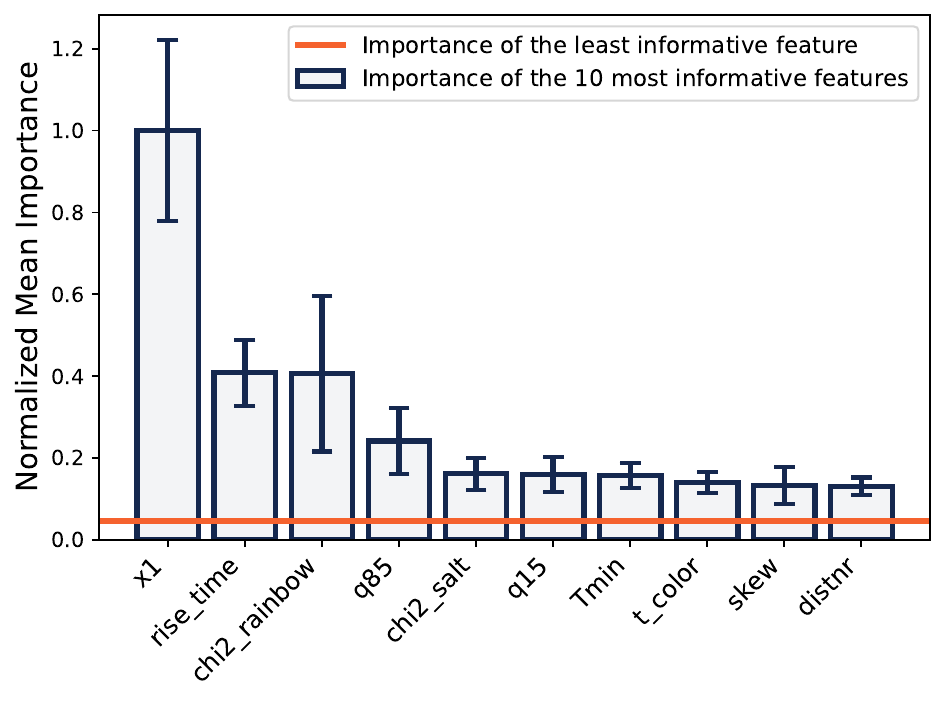}
    \caption{Normalized feature importance of the top 10 most informative feature of the classifier, average over the 100 bootstrapping iterations. The uncertainty corresponding to the standard deviation of the feature importance over the bootstrapping procedure.}
    \label{fig:feature_importance}
\end{figure}

Finally, Figure \ref{fig:feature_importance} presents the feature importance analysis. It shows the 10 most informative features, thereby providing insight into the classifier’s decision-making process. By far the most influential feature is $x_1$, the stretch parameter of the SALT2 model. This dominance reflects the dataset imbalance, with SNe Ia being by far the most common unique sources. Nevertheless, $x_1$, together with the $\chi^2$ of the SALT2 fit (5th feature), highlights the effectiveness of SALT2 in rapidly identifying SN Ia light curves. The Rainbow fit also contributes essential information to the classification, particularly through the $rise\_time$ and the fit $\chi^2$ (2nd and 3rd features), but also to a lesser extent with the temperature and its evolution (7th and 8th features).

\section{Fink implementation}
\label{sec:fink}

\textsc{NOMAI} is efficient at identifying SLSNe in the training dataset. However, for it to be useful for the scientific community, it should be able to run continuously on the ZTF data stream, and produce public classifications that can be easily retrieved. In order to meet these requirements, we incorporate the classifier into the Fink ecosystem in the form of a science module. Science modules are pipelines, built directly by the scientific community, able to process ZTF alerts and enrich them with extra information. In our case, the SLSN classification pipeline can ingest the data stream nightly and add \textsc{NOMAI}'s score as extra information to the alerts. The module is integrated into a distributed computing infrastructure relying on Apache Spark \citep{spark} to efficiently perform all steps (transient filtering, light curve aggregation, feature extraction, and classification) on distributed chunks of data coming in real-time. The complete implementation is publicly available on Fink\footnote{\url{https://github.com/astrolabsoftware/fink-science/tree/master/fink_science/ztf/superluminous}}. It should be stated that the performance scores on the training sample may not directly translate to the live ZTF stream. The training set, constructed from labeled events, is inherently biased toward specific classes and brighter sources, and therefore does not fully represent the statistical properties of the raw alert stream. Nevertheless, because the classifier relies on physically motivated features, its learned representation is expected to remain largely transferable, even under different data distributions.

\begin{figure}
    \centering
    \includegraphics[width=1\linewidth]{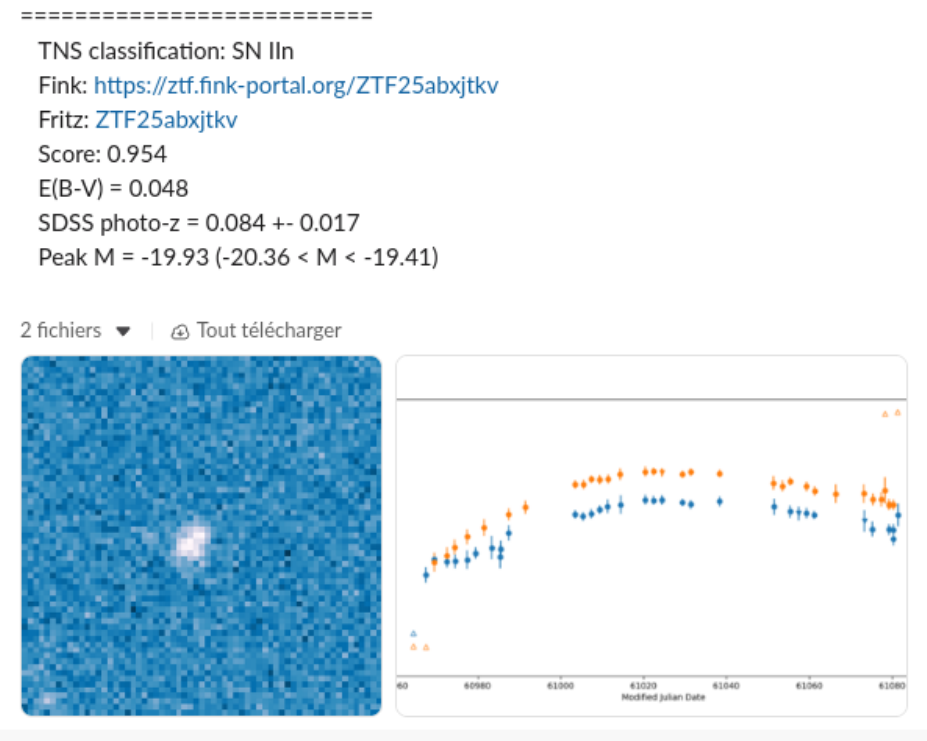}
    \caption{Screenshot of the Slack bot nightly reporting SLSN candidates. It includes, from top to bottom: the TNS classification; a link to the Fink webpage (public); a link to the Fritz webpage (private); the SLSN score from \textsc{NOMAI}; the Milky Way extinction; the SDSS photometric redshift, if available; the peak absolute magnitude based on the SDSS photometric redshift, if available; the alert stamp; and the full ZTF light curve of the source.}
    \label{fig:bot}
\end{figure}

Accessing \textsc{NOMAI}'s output directly from the ZTF alert stream can be non-trivial for users. To facilitate rapid and convenient access, we implemented a Slack bot that automatically posts all alerts classified as SLSN candidates each night to dedicated channels. As shown in Figure \ref{fig:bot}, candidates are reported with multiple contextual information, facilitating fast decision making for experts looking to the sub-stream daily.

\textsc{NOMAI} has been running live since 18/12/2025, and in what follows we discuss its real-time performance up to 18/02/2026, corresponding to two months of operation. During this period, minor corrections were applied, leading to slight adjustments toward the final version presented in this paper. However, these modifications only marginally improved the performance metrics on the training sample, and we therefore consider the results to be consistent throughout this interval.

\begin{figure}
    \centering
    \includegraphics[width=1\linewidth]{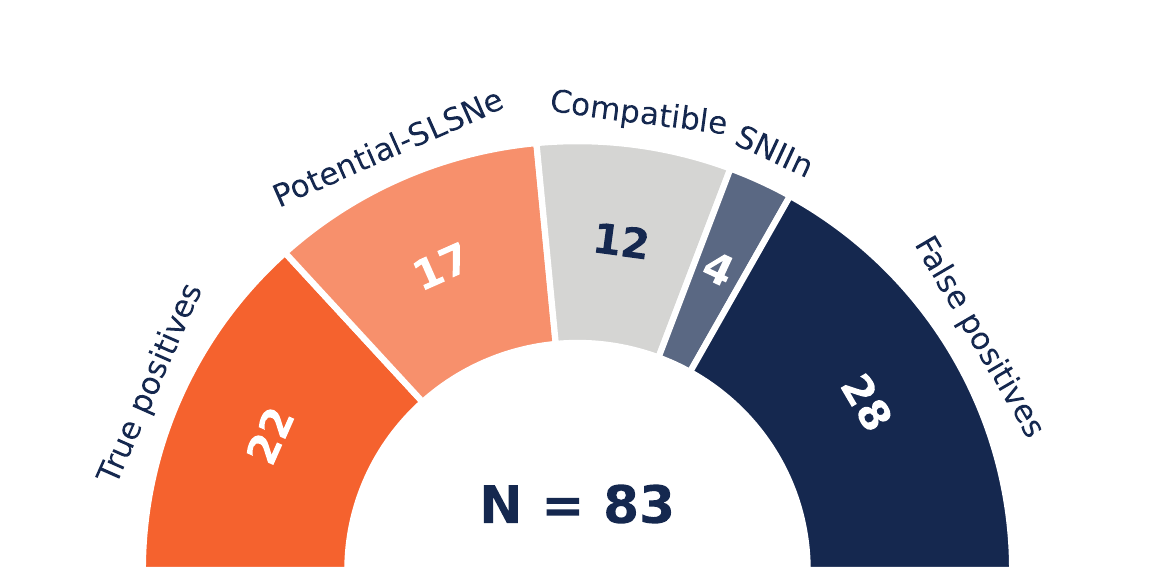}
    \caption{Classification of the 83 SLSN candidates reported by the model during two months of live operation. Candidates are divided into categories based on TNS classifications (True positives, SNe IIn, False positives) and photometric inspection (Potential-SLSNe, Compatible, False positives). The names of the sources are detailed in Appendix (Table \ref{tab:inspection})}
    \label{fig:slsn_candidates}
\end{figure}

Evaluating performance metrics on the live alert stream is challenging, as most sources remain unclassified. Despite this limitation, we attempt to provide informative indicators to assess the classifier’s behavior. To estimate the model purity, we collect all sources classified by the model as SLSNe during the two months of operation. A total of 83 unique sources were identified, among which 55 have received a TNS classification. For the remaining objects, we inspect their photometry and adopt a simple classification scheme: “True positives” correspond to TNS-classified SLSNe, or CCSNe exceeding the absolute-magnitude threshold; “Potential-SLSN” are sources displaying clear SLSN characteristics, such as long duration, blue color, a faint host, or a high absolute magnitude inferred from the photometric redshift; “Compatible” refers to transient sources that could be SLSNe but do not exhibit most of their typical features; and “False positives” correspond to non-SLSN TNS-classified sources, or events for which the photometry alone clearly rules out the SLSN scenario. Figure \ref{fig:slsn_candidates} summarizes the classification of the 83 candidates. Although a definitive purity cannot be established, the results broadly match the performance observed during the training phase. A total of 39 sources can be considered correct identifications, including 22 confirmed in TNS. However, a significant fraction of misclassifications remains, with 32 false positives. These mainly correspond to AGNs or cases of poor photometry leading to artificially flat light curves. The model also identified four SNe IIn which, although they do not reach the absolute magnitude threshold, remain of interest for the SLSN community, as they could provide insight into the possible continuum between SNe IIn and SLSNe-II.

We also attempt to estimate the completeness of \textsc{NOMAI} by collecting all active TNS-classified sources that triggered a detection during the two-month period (excluding events used in the training sample). After verifying that their light curves satisfy the minimal requirements of the classifier (Sections \ref{subsec:transient} and \ref{subsec:quality}), a total of 279 labeled sources were gathered. Among these, 24 objects were classified as SLSNe, or as CCSNe with absolute magnitudes exceeding the adopted threshold. As shown in Figure \ref{fig:slsn_candidates}, the classifier correctly identifies 22 (92\%) of them, suggesting a high level of completeness. However, as indicated by the “potential-SLSN” candidates, TNS-classified events likely represent only a fraction of all SLSNe present in the stream, and it is impossible to determine how many may have been missed by human classification. Consequently, this value should be regarded as an upper limit on the classifier’s completeness.

\section{Discussion}
\label{sec:discussion}

\begin{figure*}
    \centering
    \includegraphics[width=0.85\linewidth]{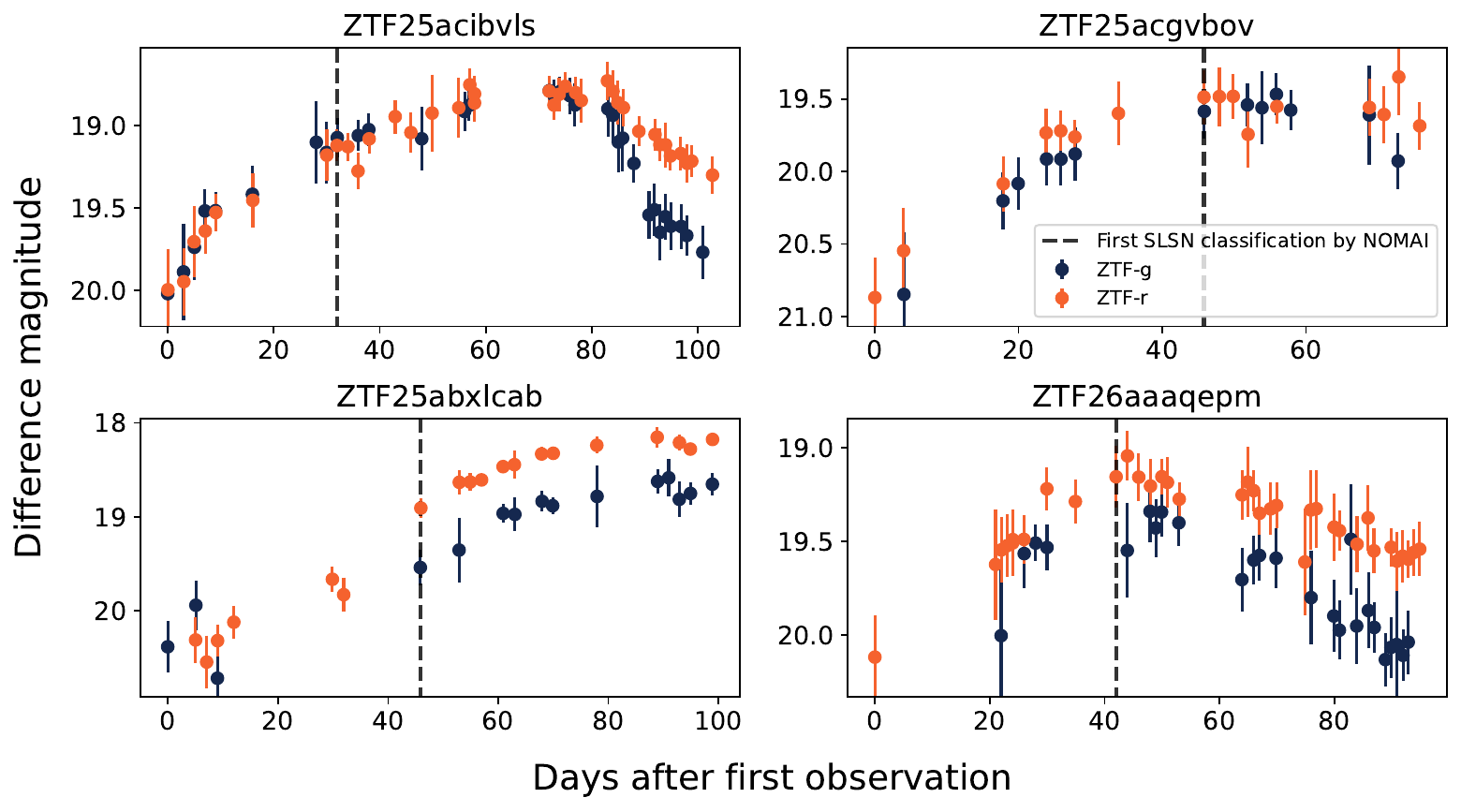}
    \caption{Examples SLSN candidate light curves identified by \textsc{NOMAI}. The two top panels and the bottom-left panel show sources that were flagged as promising candidates prior to attracting expert attention and were later spectroscopically confirmed as SLSNe. The bottom-right panel shows a promising candidate identified and shared by \textsc{NOMAI} that did not receive spectroscopic follow-up. The dashed vertical line indicates the first alert classified as SLSN by \textsc{NOMAI}.}
    \label{fig:examples}
\end{figure*}

\textsc{NOMAI} has demonstrated robust performance both on the training dataset and under real-time operating conditions. It has been running continuously since 18/12/2025 and is already used by experts as a recommendation system for further analysis. The daily scans and automated summaries have contributed to the identification of previously unclassified SLSN candidates. Some examples are shown in Figure \ref{fig:examples}, where several sources were flagged by \textsc{NOMAI} as promising SLSN candidates prior to attracting expert attention and before their eventual spectroscopic confirmation, illustrating its ability to spread noteworthy events at early stages. The figure also includes a compelling candidate (bottom-right panel) that, despite being identified and shared by NOMAI, did not receive spectroscopic follow-up, highlighting a limitation in current workflows where promising events can be missed. Such cases are common and emphasize the importance of classification tools in prioritizing and increasing the visibility of candidates that might otherwise go unnoticed.

While the current methodology provides a solid foundation, it also presents limitations that could be improved in the future. In this section we discuss these limitations and possible developments, and we compare it to the current landscape of existing SLSN classifiers.

\subsection{Limitation and future development}
\label{subsec:lim_and_future}

One of the main limitations of the present approach concerns the minimum light-curve duration required for reliable classification. In the current implementation, the classifier uses the first 30 days of photometric evolution, which is sufficient to achieve the high completeness reported in this work and therefore does not represent a fundamental limitation of the method itself. In practice, however, earlier classifications would be desirable for survey operations. Rapid identification is important to trigger follow-up observations, in particular spectroscopy obtained close to maximum light. Even if SLSNe are generally long lived sources, requiring a full 30-day baseline may reduce the fraction of objects that can be spectroscopically observed during their most informative phases. This requirement also potentially biases the classifier towards slowly evolving SLSNe, neglecting faster ones. Future work could focus on reducing this minimum duration while maintaining satisfactory performances.

A second limitation arises from the limited use of host-galaxy information in the current feature set. The classifier relies primarily on the transient light curve and only indirectly includes environmental information through the alert feature describing the angular distance to the nearest reference object, and the SDSS photometric redshift if available. While these numbers are very informative, they provide no information about the host properties, meaning that morphology, brightness, or local environment are not exploited. Incorporating image-level information \citep[as shown in][]{pure_host_classification, needle} or host-galaxy features \citep[such as][]{fleet} derived directly from survey images could therefore provide additional discriminatory power and improve classification performance.

These developments will be particularly relevant to process the LSST public alert stream, which delivers an unprecedented volume of transient detections. In practice, adapting \textsc{NOMAI} from ZTF to LSST should remain straightforward from a technical perspective, since the feature extraction relies on light-curve fits that require few observations and uses Rainbow, which is intrinsically designed to handle multi-band photometry. However, the main challenge will lie in constructing an informative and representative training dataset.

\subsection{Comparison to other classifiers}
\label{subsec:compare}

As discussed in the introduction, \textsc{NOMAI} complements the various SLSN classification methods available in the literature. It is therefore natural to compare its performance with similar implementations, such as NEEDLE or FLEET, which both aim to identify SLSNe directly from the ZTF data stream without relying on spectroscopic redshifts. In this context, FLEET 2.0 \citep{fleet2}, the most recent version of the model, reports approximately 50 (40)\% purity and 50 (30)\% completeness when classifying light curves at late (early) phases using a classification threshold of P(SLSN)>0.5. These values are broadly similar to the metrics obtained in this work, namely 58\% purity and 66\% completeness. In contrast, NEEDLE \citep{needle} reports higher performance, with an average purity of 84\% and completeness of 75\%, suggesting that the detection and reference image cutouts are highly informative for classification. However, it remains difficult to determine to what extent these differences reflect intrinsically better performance or simply variations in the evaluation conditions.

More generally, direct comparisons between different classifiers should be interpreted with a lot of caution. Reported performance metrics depend strongly on the characteristics of the datasets used for training and evaluation, including the selection criteria, preprocessing steps, class distributions, and the survey from which the data are drawn. Consequently, without a dedicated benchmarking framework, any strict quantitative comparison between models is impossible. At the same time, the diversity of methodologies used by existing classifiers suggests that they may capture complementary information from the data. In this context, maintaining multiple independent classification approaches may be beneficial, as combining their outputs could ultimately provide a more robust identification of SLSN candidates.

\section{Conclusion}
\label{sec:conclusion}

In this work, we presented the methodology of \textsc{NOMAI}, running on the Fink broker. It relies on a strong filtering step that removes most non-transient events, allowing the efficient retrieval of full light curves for the sources. These light curves are subsequently processed through feature extraction using simple statistical quantities as well as model-fitting procedures based on SALT2 and Rainbow. These features are highly informative and summarize the inhomogeneous light curves into fixed-size arrays.

The training set is composed of 5280 unique sources, among which 225 are SLSNe, carefully labeled by combining public classifications from TNS and private labels from Fritz, along with a reclassification of some sources based on their absolute magnitudes. After converting the dataset to alert format, an XGBoost classifier is trained on these features to separate SLSNe from the rest of the ZTF sources. It reaches 66\% completeness and 58\% purity on the training sample, making it, a priori, a powerful tool for analyzing the ZTF alert stream. However, these metrics are only valid for the training set, which, by the nature of the labeling process, is subject to various biases.

\textsc{NOMAI} has been integrated into the Fink ecosystem and has been processing public alerts every night for the past months. We therefore attempt to estimate its real-world performance by evaluating two months of SLSN classifications. Although exact metrics cannot be computed because of the lack of labels, the purity predicted from the training sample appears broadly consistent with the results obtained under real conditions. Regarding completeness, the classifier performs very well, successfully identifying 22 of the 24 active TNS SLSNe during this period. This value, however, does not represent a true completeness, since the total number of SLSNe that triggered alerts during this time interval is unknown; it should therefore be regarded as an upper limit.

Beyond these quantitative metrics, the classifier has been continuously operating within the Fink broker and providing nightly recommendations to the community. The results are automatically shared through dedicated public Slack channels, allowing experts to quickly inspect promising candidates without the need to interact directly with the alert stream. This user-friendly interface has proven effective for scanning the large volume of incoming alerts and highlighting potentially interesting SLSN candidates. In practice, \textsc{NOMAI} has already facilitated the identification and discussion of several events within the community, demonstrating its usefulness as a real-time recommendation tool for transient follow-up.

Many questions remain about SLSNe, since their progenitors, powering mechanisms, and environments are still debated. Increasing the total number of events is essential to improve our understanding of these extreme transients. Tools such as the classifier presented in this work contribute to this effort by enabling the live identification of promising SLSN candidates within large alert streams, facilitating targeted follow-up observations. This need will become even more pressing with the advent of the Legacy Survey of Space and Time conducted by the Vera C. Rubin Observatory, which is expected to discover superluminous supernovae in unprecedented numbers. In this context, automated pipelines capable of identifying these events in real time will not only be useful, but necessary for fully exploiting the scientific potential of next-generation time-domain surveys.\\\\

\textit{\textbf{Acknowledgment:} This project is funded by the European Union (ERC, project number 101042299, TransPIre). Views and opinions expressed are however those of the author(s) only and do not necessarily reflect those of the European Union or the European Research Council Executive Agency. Neither the European Union nor the granting authority can be held responsible for them. Based on observations obtained with the Samuel Oschin Telescope 48-inch and the 60-inch Telescope at the Palomar Observatory as part of the Zwicky Transient Facility project. ZTF is supported by the National Science Foundation under Grants No. AST-1440341 and AST-2034437 and a collaboration including current partners Caltech, IPAC, the Weizmann Institute of Science, the Oskar Klein Center at Stockholm University, the University of Maryland, Deutsches Elektronen-Synchrotron and Humboldt University, the TANGO Consortium of Taiwan, the University of Wisconsin at Milwaukee, Trinity College Dublin, Lawrence Livermore National Laboratories, IN2P3, University of Warwick, Ruhr University Bochum, Northwestern University and former partners the University of Washington, Los Alamos National Laboratories, and Lawrence Berkeley National Laboratories. Operations are conducted by COO, IPAC, and UW. The Gordon and Betty Moore Foundation, through both the Data-Driven Investigator Program and a dedicated grant, provided critical funding for SkyPortal.}

\bibliography{sample}

\clearpage

\section*{Appendix}

\begin{table}[h!]
\centering

\small
\begin{tabular}{p{3cm} p{13cm}}
\hline
\addlinespace[2pt]
\textbf{Class} & \textbf{ZTF objectIds} \\
\addlinespace[4pt]
\hline
\addlinespace[2pt]
True positives &
ZTF24abyaowk, ZTF25abxjtkv, ZTF25aaggkxk, ZTF25acibvls, ZTF25aauczue,
ZTF25aatmfzy, ZTF25achwsgj, ZTF25abwvblg, ZTF25acemaph, ZTF25abxlcab,
ZTF25acfybcc, ZTF25abyrwhz, ZTF25abzyzac, ZTF25acgvbov, ZTF25accmopr,
ZTF25achwzws, ZTF25abmhchv, ZTF25abxfsza, ZTF25acfbplr, ZTF25acgorke,
ZTF25acelgwl, ZTF26aaakmmc \\
\addlinespace[4pt]
\hline
\addlinespace[2pt]
Potential-SLSN &
ZTF25acelkdo, ZTF25acgtdgz, ZTF25achvfus, ZTF25acicrjr, ZTF18aaweguw,
ZTF25acjscjg, ZTF25acasutk, ZTF25acekwdl, ZTF25achctne, ZTF25acehskf,
ZTF25acgojmh, ZTF25acldhis, ZTF25acbjqvq, ZTF25achbrei, ZTF25acghjar,
ZTF25abxpson, ZTF26aaadhkv \\
\addlinespace[4pt]
\hline
\addlinespace[2pt]
Compatible &
ZTF25accfued, ZTF25aceeehm, ZTF25acendyn, ZTF25acjedts, ZTF25acgakau,
ZTF25ackkmlt, ZTF25achqjdq, ZTF25acktbjh, ZTF26aaaqepm, ZTF26aabdhlb,
ZTF25acfpqje, ZTF26aaaqdbi \\
\addlinespace[4pt]
\hline
\addlinespace[2pt]
SNIIn &
ZTF25aadhpuz, ZTF25abxqslz, ZTF25acgggyl, ZTF25acggpbd \\
\addlinespace[4pt]
\hline
\addlinespace[2pt]
False positives &
ZTF18aaiuwhq, ZTF25abubydw, ZTF25abuovxv, ZTF18aczazkn, ZTF23aagapcz,
ZTF25abxqrvf, ZTF25aahmacs, ZTF25abzivnk, ZTF25acgtjkh, ZTF24aarwhrh,
ZTF18aaxqitv, ZTF25acinfdh, ZTF25acfjqjb, ZTF25acjqzbs, ZTF25abwtoay,
ZTF25aadoqxb, ZTF18aaisaqo, ZTF25acemsej, ZTF20aahbzae, ZTF25abwbkdq,
ZTF25abgviek, ZTF25abtgdoi, ZTF25abkeuac, ZTF25aceqoyb, ZTF25acefgbd,
ZTF21acqhbdp, ZTF25ackarzf, ZTF26aaarplj \\
\addlinespace[4pt]
\hline
\end{tabular}
\caption{List of inspected ZTF objects grouped by classification outcome. Potential-SLSN were purely classified based on present public photometric data, therefore they might be confirmed or rejected in the future.}
\label{tab:inspection}

\end{table}

\clearpage
\begin{table}[h!]
    \centering
    \renewcommand{\arraystretch}{1.2}
    \begin{tabular}{l p{7cm}}
         \hline
         \addlinespace[2pt]
         \multicolumn{2}{c}{\textbf{Rainbow features}} \\
         \addlinespace[2pt]
         \hline
         \addlinespace[2pt]
         $amplitude$ & Amplitude parameter of the Bazin flux model. \\
         $rise\_time$ & Characteristic rise timescale ($\tau_{\mathrm{rise}}$) of the Bazin model. \\
         $fall\_time$ & Characteristic decay timescale ($\tau_{\mathrm{fall}}$) of the Bazin model. \\
         $T_{min}$ & Late-time temperature from the temperature model. \\
         $T_{max}$ & Initial temperature from the temperature model. \\
         $t_{color}$ & Cooling timescale. \\
         $chi2\_rainbow$ & $\chi^2$ value of the Rainbow model fit. \\
         $snr\_amplitude$ & Signal-to-noise ratio of the amplitude parameter. \\
         $snr\_rise\_time$ & Signal-to-noise ratio of the rise time parameter. \\
         $snr\_fall\_time$ & Signal-to-noise ratio of the fall time parameter. \\
         $snr\_T_{min}$ & Signal-to-noise ratio of the minimum temperature. \\
         $snr\_T_{max}$ & Signal-to-noise ratio of the maximum temperature. \\
         $snr\_t_{color}$ & Signal-to-noise ratio of the cooling timescale. \\
         \addlinespace[4pt]
         \hline
         \addlinespace[2pt]
         
         \multicolumn{2}{c}{\textbf{SALT2 features}} \\
         \addlinespace[2pt]
         \hline
         \addlinespace[2pt]
         $x_0$ & Amplitude parameter of the SALT2 model. \\
         $x_1$ & Light-curve stretch parameter from the SALT2 model. \\
         $c$ & Color parameter from the SALT2 model. \\
         $z$ & Redshift estimated from the SALT2 fit. \\
         $chi2\_salt$ & $\chi^2$ value of the SALT2 model fit. \\
         \addlinespace[4pt]
         \hline
         \addlinespace[2pt]
         
         \multicolumn{2}{c}{\textbf{Statistical / Alert features}} \\
         \addlinespace[2pt]
         \hline
         \addlinespace[2pt]
         $flux\_amplitude$ & Difference between the maximum and minimum flux. \\
         $max\_slope$ & Maximum flux slope between two consecutive observations. \\
         $std\_flux$ & Standard deviation of the flux. \\
         $duration$ & Time difference between the last and the first observation. \\
         $q15$ & 15th percentile of the observation epochs. \\
         $q85$ & 85th percentile of the observation epochs. \\
         $skew$ & Skewness of the flux distribution, measuring asymmetry. \\
         $distnr$ & Mean angular distance to the nearest object in the reference image. Available in ZTF alert packets. \\
         $E(B-V)$ & Milky Way extinction estimated using \texttt{SkyCoord} from \texttt{astropy}. \\
         \addlinespace[4pt]
         \hline
         
    \end{tabular}
    \caption{Summary of all features used by the model, grouped into Rainbow, SALT2, and statistical/alert features.}
    \label{tab:all_features}

\end{table}
\end{document}